\documentclass{cpcauth}

\usepackage{epsfig}

\begin{document}
\begin{frontmatter}
\title{Charge Dependence of Temperature-Driven Phase Transitions of
Molecular Nanoclusters: Molecular Dynamics Simulation}

\author{S. Pisov and A. Proykova}
\address{University of Sofia, Faculty of Physics, 5 James Bourchier Blvd.,
Sofia-1126, Bulgaria}

\begin{abstract}
Phase transitions (liquid-solid, solid-solid) triggered by temperature changes
are studied in free nanosized clusters of $TeF_6$ ($SF_6$) with different
negative charges assigned to the fluorine atoms. Molecular dynamics
simulations at constant energy show that the charge increase from $q_F$=0.1$e$
 to
$q_F$=0.25$e$ shifts the melting temperature towards higher values and 
some of the metastable solid states disappear. The increased repulsive
 interaction maintains the order in molecular systems at higher temperatures.
\end{abstract}

\begin{keyword}
molecular dynamics \sep quenching \sep phase transition
\PACS 36.40Ei \sep 64.70Kb \sep 61.50-f
\end{keyword}

\end{frontmatter}
\section{Introduction}
Small free clusters of atoms or molecules exhibit solid-like or
liquid-like properties that differ from the properties of their bulk
counterpart. It has been realized that the cluster structure
(microcrystalline or amorphous) \cite{Ref-1,Ref-2} depends on the
production method. Clusters made of $AF_6$ molecules ($A = S,Se, Te, U, W$)
have already been examined in some detail by several groups 
\cite{Ref-2,Ref-3,Ref-4,Ref-5,Ref-6}
 to establish the existence of various structures and
transformations between them. Some of the microcrystalline states
 coexist dynamically 
\cite{Ref-5,Ref-6} in a given temperature interval and those which are only
 locally stable phases disappear when the cluster size increases as 
was confirmed both experimentally \cite{Ref-1} and theoretically
\cite{Ref-6}. Those states correspond to a partial order of the molecular
axes of symmetry. The  system becomes completely orientationally 
ordered at very low temperatures. The transition rate between the 
 ordered and disordered states can be
retrieved from
the potential energy surface (PES)
of the system, \cite{Ref-7}. It has been shown in \cite{Ref-6} that
clusters of the same numbers of $TeF_6$ or $SF_6$ molecules
exhibit different dynamics despite the same symmetry 
of the molecules. The reason is that the topography of PES in the case of
$SF_6$ clusters is shallower than that of $TeF_6$ clusters.

In the present study we explore
the changes of the cluster PES due to the changes of the charge
distribution in a single molecule. Our hypothesis is that the molecular
polarization is changed by using different production methods. In
order to find out the influence of the charge changes on
orientational order-disorder phase transitions we have simulated the
temperature behavior of molecular clusters with the help of a constant
energy molecular dynamics.

\section{Interaction Potential and Computational Procedure}
The main feature of the intermolecular interaction is the dependence on the
mutual orientations of the molecules. There are experimental indications
that $AF_6$ molecules can be considered as rigid octahedra to a reasonable
extent. A small negative charge should  be assigned to the fluorine
atoms to account for the chemical bond \cite{Ref-3}.

 The intermolecular potential is presented as a sum of
atom-atom interaction (fluorine-fluorine, tellurium-tellurium, 
fluorine-tellurium) to account for the orientational anisotropy:

\begin{equation}
\label{potential}
\begin{array}{l}
U_{pw}(i, j) = \sum\limits_{\alpha, \beta = 1}^7
\Biggl \lbrack
4 \, \epsilon_{\alpha \beta}\biggl \lbrack
\biggl(
\frac{\sigma_{\alpha \beta}}{r_{ij}^{\alpha \beta}}
\biggr)^{12}
-
\biggl(
\frac{\sigma_{\alpha \beta}}{r_{ij}^{\alpha \beta}}
\biggr)^6
\biggr \rbrack \\
+\frac{q_{i \alpha} q_{j \beta}}{4 \pi \epsilon_0 r_{ij}^{\alpha \beta}}
\Biggr \rbrack
\\

\\
U_{pot} = \sum\limits_{
i,j = 1 \,(i < j)
}^n
{
U_{pw}(i, j)
}
\end{array}
\end{equation}
\noindent
where $\alpha,\beta$ denote either a fluorine or a tellurium
(sulfur) atom; $r^{\alpha\beta}_{ij}$ is the distance between an
$\alpha$-atom in the $i-th$ molecule and a $\beta$-atom in the $i-th$
molecule; $n$ is the number of the molecules in the cluster. 
The parameters $\sigma_{\alpha \beta}$ and $\epsilon_{\alpha \beta}$   
have been fitted to the experimental diffraction results, \cite{Ref-2}.

The Coulomb term accounts for the small negative charge $q_F$
assigned to the fluorine atoms and the positive charge carried
by the central tellurium atom, $q_{te}$ = 6 $q_F$, which
ensures a neutral molecule at distances much larger than the
molecular size. Here we compare the temperature-driven
transitions for the case of $q_F = 0.1 e$ and $q_F = 0.25 e$, where e
is the electron charge.

The charge $q_F$ has been computed using LCAO (linear combinations of
atomic orbitals) with planar basic functions ($q_F = 0.1 e$) and Gaussian
basic functions ($q_F = 0.25 e$).

The potential, Eq.\ref{potential}, has been used to solve the classical
equation of motion written in the Hamiltonian form with the help of a
constant-energy MD method. The velocity-Verlet \cite{Ref-8} algorithm
with a time step of $5 fs$ has been implemented. This is a step optimized
in \cite{Ref-6} to satisfy the requirements for long MD runs
necessary to detect phase changes in a computer experiment. The heating
(cooling) of the system is performed by velocity rescaling and
consequent thermalization \cite{Ref-3}.

\section{Results and Conclusions}
\begin{figure}[h]
\mbox{\epsfig{file=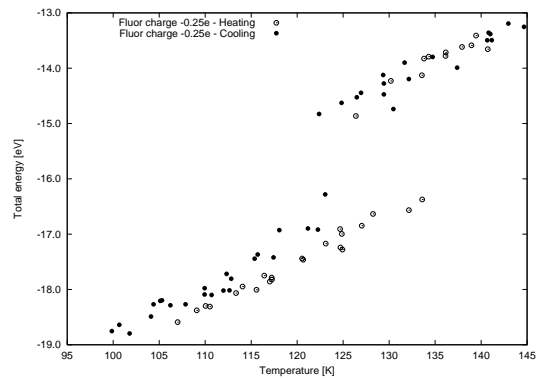,height=50mm}}
\caption{Caloric curve for $TeF_6$(89) cluster, with $q_F = 0.25 e$}
\label{Fig-1}
\end{figure}

\begin{figure}[h]
\mbox{\epsfig{file=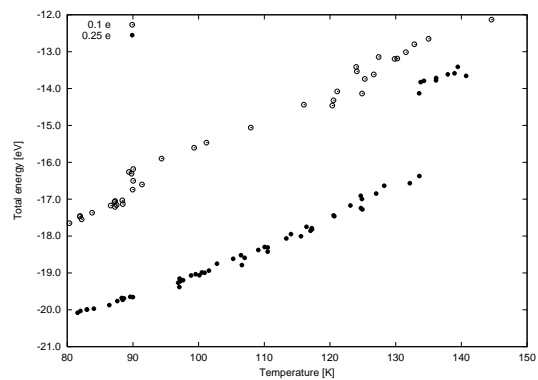,height=50mm}}
\caption{Caloric curve for $TeF_6$(89) cluster, with $q_F = 0.1 e$ and
$q_F = 0.25 e$}
\label{Fig-2}
\end{figure}

\begin{figure}[h]
\mbox{\epsfig{file=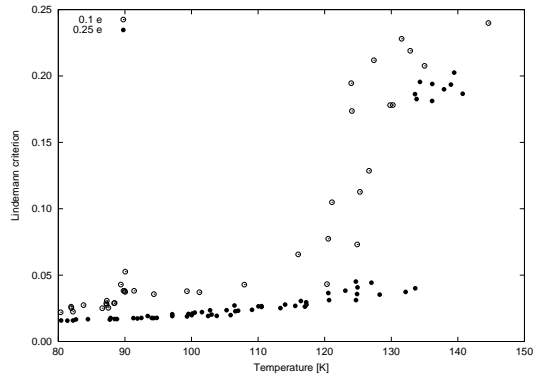,height=50mm}}
\caption{Lindemann criterion for $TeF_6$(89) cluster, with $q_F = 0.1 e$ 
and $q_F = 0.25 e$}
\label{Fig-3}
\end{figure}

We have investigated the thermal behavior of clusters consisting of 89
$TeF_6$ ($SF_6$) molecules in the temperature interval (50 $\div$  140
K). The clusters of $TeF_6$ molecules melt above 125K ($q_F = 0.1 e$) and
130K ($q_F = 0.25 e$) as is seen from the caloric curves in
Fig.\ref{Fig-1}, Fig.\ref{Fig-2} and the Lindemann criterion,
Fig.\ref{Fig-3}. The Lindemann coefficient $\delta < 0.08$ corresponds
to a solid-like phase ($\delta < 0.1$ for bulk 
systems) \cite{Ref-11}. Melting is a discontinuous transition even in the
case of nanosized clusters. The heating and cooling of clusters
(Fig.\ref{Fig-1}) demonstrate a hysteresis, which signals a discontinuous
transition. The charge increase shifts the melting point towards higher
temperature: in comparison to a cluster with less charged atoms,
Fig.\ref{Fig-2}. [ the larger charge, the more robust is the cluster 
].

The freezing point shifts towards lower temperatures.

\begin{figure}[h]
\mbox{\epsfig{file=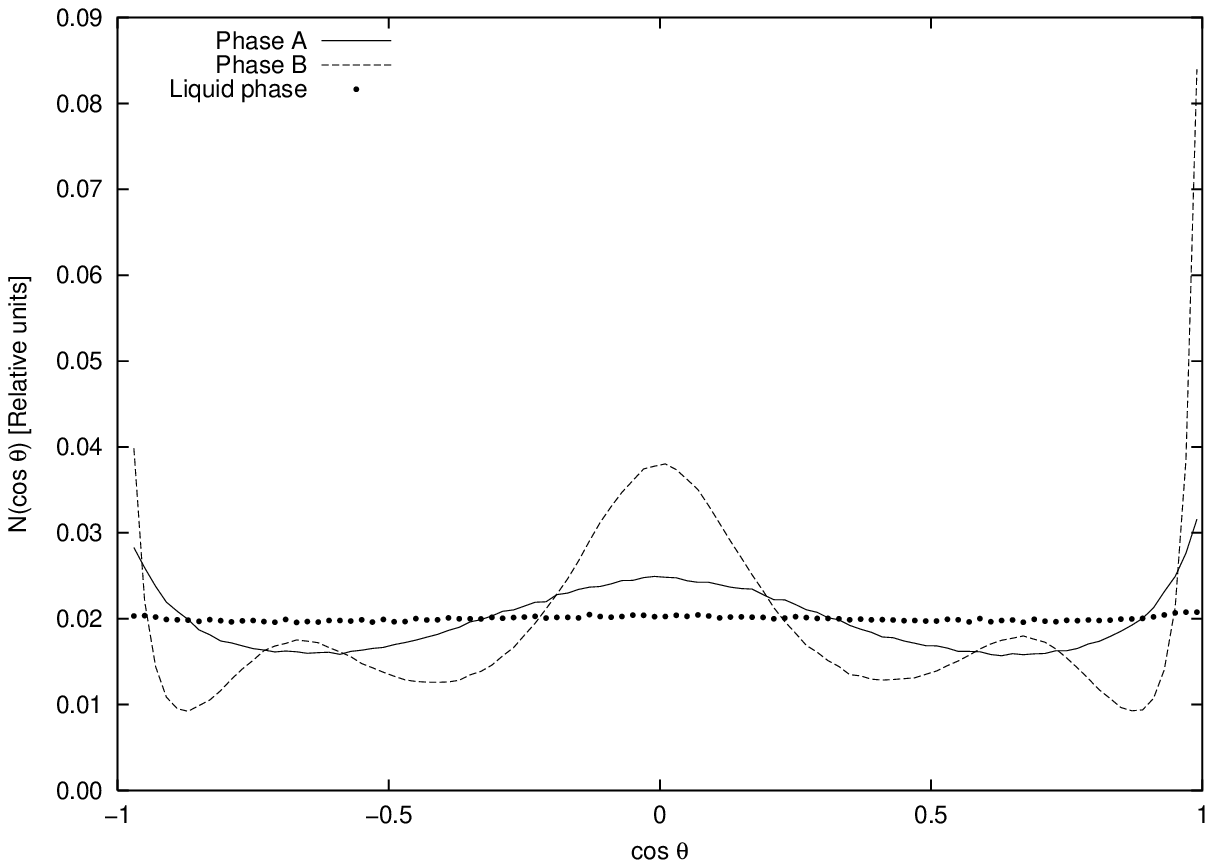,height=50mm}}
\caption{Orientation distribution $N(cos \theta)$ for $TeF_6$(89) cluster, 
with $q_F = 0.1 e$}
\label{Fig-4}
\end{figure}

\begin{figure}[h]
\mbox{\epsfig{file=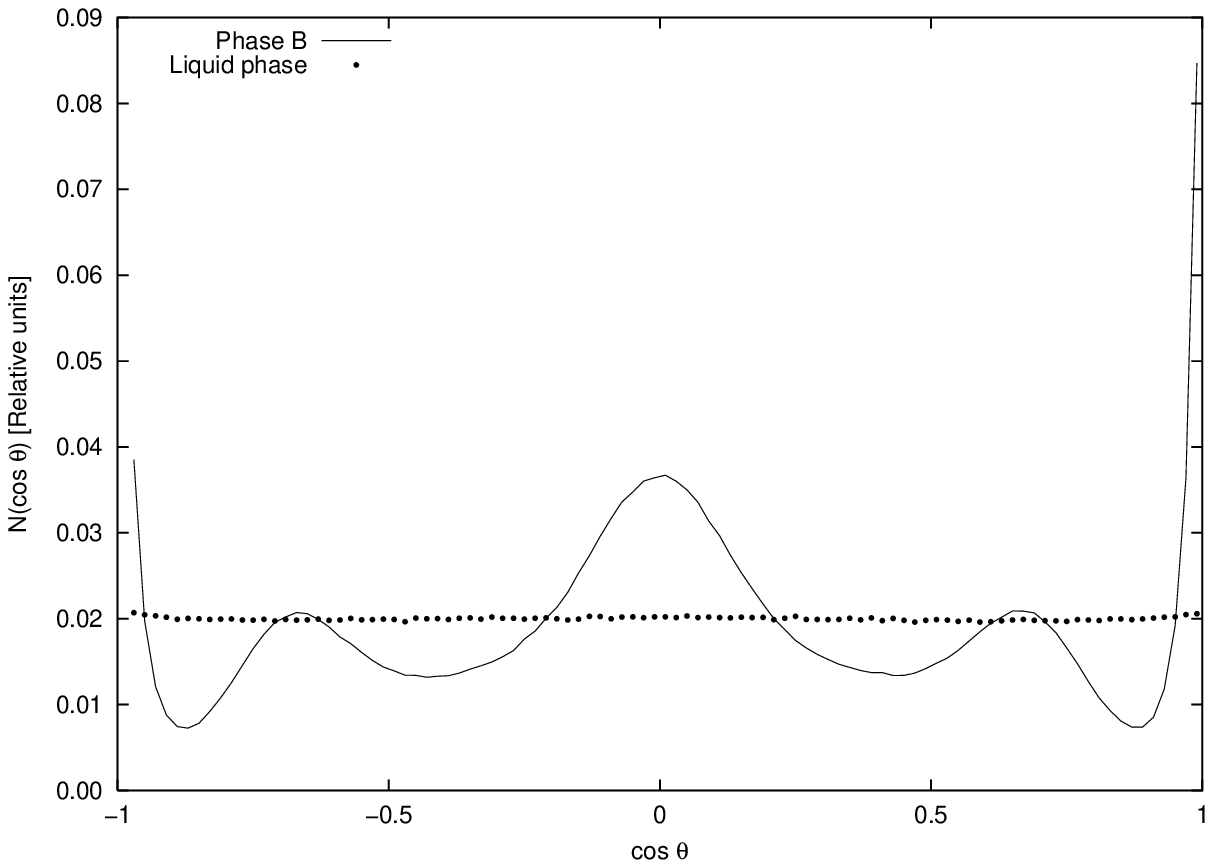,height=50mm}}
\caption{Orientation distribution $N(cos \theta)$ for $TeF_6$(89) cluster, 
with $q_F = 0.25 e$}
\label{Fig-5}
\end{figure}

\begin{figure}[h]
\mbox{\epsfig{file=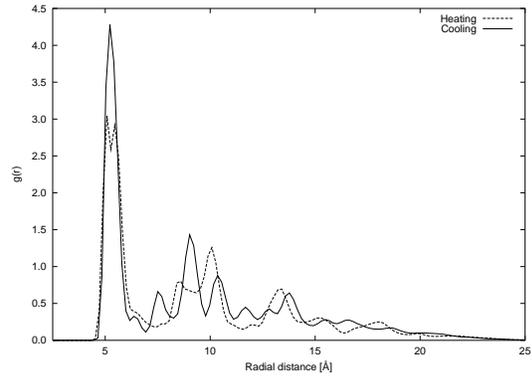,height=50mm}}
\caption{Radial distribution g(r) for $TeF_6$(89) cluster, with $q_F =
0.25 e$}
\label{Fig-6}
\end{figure}

\begin{figure}[h]
\begin{tabular}{c}
\mbox{\epsfig{file=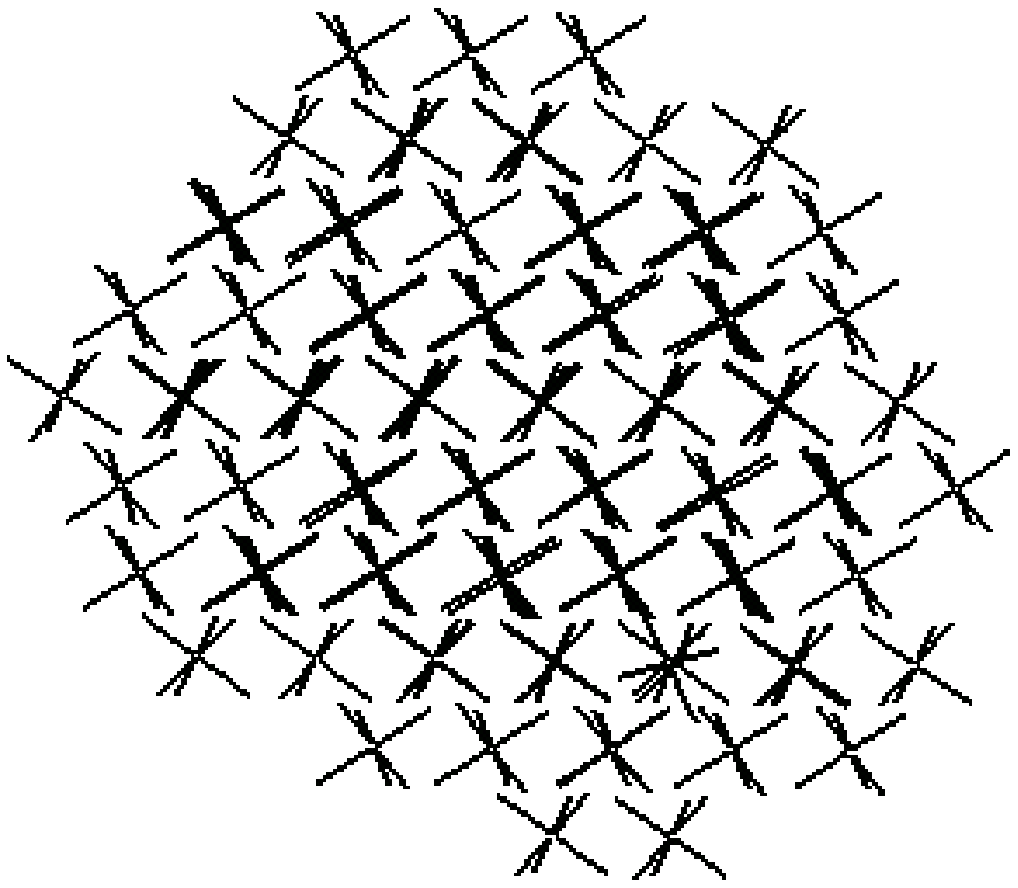,height=50mm}} (a)\\
\mbox{\epsfig{file=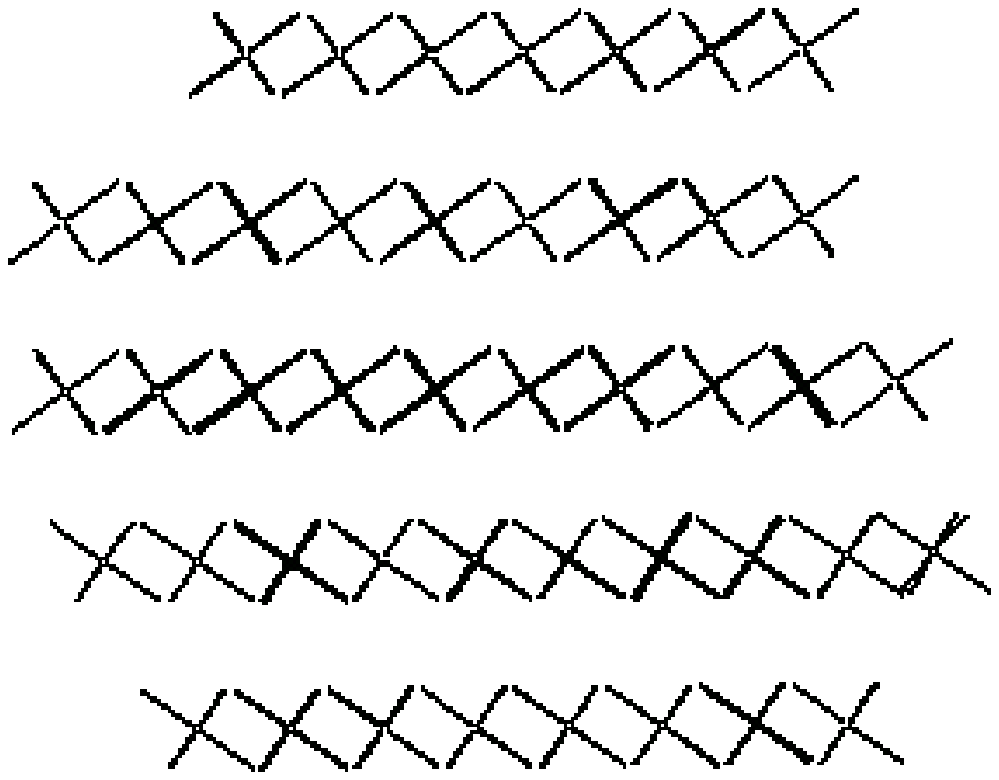,height=50mm}} (b)
\end{tabular}
\caption{(a) A quenched configuration of $TeF_6$(89) clusters with $q_F
= 0.1 e$: a sequence of two-one-two layers with specifically
oriented molecules is seen; (b) A quenched configuration of the same
cluster with $q_F = 0.25 e$. The cluster crystallizes in another
structure: three rows in one orientation and two rows in another
orientation.}
\label{Fig-7}
\end{figure}

The hysteresis area is larger for the case of $q_F = 0.25 e$ than for
$q_F = 0.1 e$. One could speculate about a "larger memory" in systems
having a "larger charge". Another important distinction between less and 
more charged fluorine atoms is the structure adopted by the clusters
below the freezing point. The distributions $N(cos\theta)$ of the mutual
molecular orientations with $q_F = 0.1 e$, Fig.\ref{Fig-4}, show
that they transform step by step from liquid to a partially ordered solid
(phase A) to an ordered solid (phase B). The clusters with $q_F = 0.25 e$,
Fig.\ref{Fig-5}, transform directly from liquid to an ordered state. The
radial distribution g(r) \cite{Ref-10}, Fig.\ref{Fig-6}, is a diagnostic
of the lattice structure adopted by clusters.
We compute g(r) for the molecular center of masses which is insensitive to
the molecular orientations. Fig.\ref{Fig-6} was obtained as 
follows: starting from a low-temperature configuration obtained from 
simulations of a cluster with $q_F = 0.1 e$, we change to $q_F = 0.25 e$
and heat the cluster until it melts. Then the cluster is cooled to a solid
state. g(r) are plotted at the same $T = 110K$ on the cooling and heating
branch. The structures are obviously different. Typical configurations
obtained by quenching \cite{Ref-9} of the MD trajectory are shown in
Fig.\ref{Fig-7} [(a) - for $q_F = 0.1 e$; (b) - for $q_F = 0.25 e$]. In
both cases we obtain two orientations populated by the molecules but
the arrangements of "layers" in the clusters are different.

We conclude that the charge change influences both the melting temperature
and structural transformations. Meta-stable solid states with partial order
disappear for the case of larger charge. The volume of the cluster
decreases when the charge increases and the cluster is more robust on heating.

Our results show that the charge increase shifts the transitions
temperature towards higher values and some of the metastable states
disappear. This confirms our understanding that the repulsive interaction
maintains the order in molecular systems.

{\bf Acknowledgments}

NATO Grant (CLG SA(PST.CLG.976363)5437 is acknowledged.
The work has been partially supported by the University of Sofia
Scientific Fund (2001).

%
%

\end{document}